# Resource Allocation for Secure Ultra-Reliable Low-Latency-Communication in IoT Applications

S. Sorkhi Asbaghi, M. Mohassel Feghhi, J. Musevi niya

*Faculty of Electrical and computer Engineering, University of Tabriz, Tabriz, Iran.*
*mohasselfeghhi@tabrizu.ac.ir. sorkhisolmaz@gmail.com. niya@tabrizu.ac.ir*

*Abstract*- **The Internet of Things (IoT) has a significant demand in society due to its features, and it is constantly improving. In the context of wireless technology, Ultra-reliable and low-latency communication (URLLC) is one of the essential and challenging services in fifth-generation (5G) networks and beyond. The research on URLLC is still in its early stages due to its conflicting requirements, regarding high reliability and low latency. In this paper, we study the performance of secure short-packet communications and resource allocation in IoT systems. To this end, we investigate a health center automation, where the goal of the access point is to send critical messages to devices without eavesdropping. In this context, our goal is to maximize the weighted sum throughput and minimize the total transmit power, respectively. The problems of maximizing the weighted sum throughput, and minimizing the total transmit power are non-convex and hard to solve. To overcome this challenge, we use efficient mathematical techniques, such as the block coordinate descent (BCD) method and gradient ascent algorithm; we also use estimation methods such as Ralston, Heun, and forward-backward, in the derivative part of the gradient ascent algorithm. The simulation results show the performance advantages of the BCD algorithm and the gradient ascent in the short packet transmission scheme, also the simulation results show the superiority of the proposed methods in most cases.**

*Index Terms*- Resource Allocation, Ultra-Reliable Low Latency Communications, Internet of Things, Short Packet communications.

I. INTRODUCTION

Internet of things (IoT) means a set of objects and equipment that communicate with each other by connecting to a network. Actually the IoT has made it possible for everything around us to be connected. IoT-based services have grown significantly in recent years, and because of that, many articles have been presented to examine the IoT and its challenges. In [1] and [2], the authors have studied the IoT and its architecture, and this article also discusses some of the future ideas of the IoT, also in [2] the known threats to the IoT are examined in different layers, and a set of security guidelines is also provided. .In [3] and [4], the authors have analyzed the security of the IoT, its current threats, and vulnerabilities; they have also emphasized the importance of having a platform to secure the IoT environment.

Wireless communications with the ability to connect to the internet are expanding very fast. Therefore 5G technology has become one of the exciting topics in wireless research. In [5] the architecture of the

5G IoT is presented, the layers of the IoT are introduced; together with describing some important research areas in the field of IoT. In [6] and [7], the 5G and the 6G radio developments and active technologies for the IoTs are introduced, also the IoT application requirements and the standard infrastructures of the IoT, including the eMTC and the NB-IoT are studied. In [8] Orthogonal frequency-fivision multiplexing (OFDM) based 5G radio interface has been studied; Here two types of factory automation with different dimensions are considered, in which the sub-millisecond radio transmission can be guaranteed. This paper discusses the signal-to-noise ratio (SNR), the number of antennas and the signal bandwidth as the performance metrics.

Security is an important issue that has received attention with the widespread use of the internet. Since wireless communications are broadcast in nature, IoT communication systems are at risk in terms of security. Indeed it is one of the disadvantages of wireless communication, which must be resolved to receive data successfully. Paper [9] examines the performance of short packet communication in the IoT system with the presence of a multi-antenna eavesdropper. In [10], to combat eavesdropping in downlink networks, secure transmission schemes are designed for non-orthogonal multiple access (NOMA) networks. The goal of 5G communication and beyond is to support three important types of connectivity: enhanced Mobile Broadband (eMBB), massive Machine Type Communication (mMTC), and Ultra-Reliable Low-Latency Communication (URLLC) [11].

In paper [12], short packet communication is investigated, and it is investigated which protocols are suitable for transmitting these packets; then, the Ultra-Reliable Communication (URC) design challenges are considered. The URC is used in applications, where require low latency, such as machine-to-machine communication. Obviously, ultra-reliability is involved with the need for low latency; this has made the URLLC very important and challenging [13]. In [14], recent developments of URLLC are studied, and this paper focuses on a wide range of techniques and methods related to URLLC requirements. Ultra-reliable low-latency communications are essential in critical applications such as factory automation, e-health, autonomous driving, etc., because all these applications require very low latency, and negligible packet error probability [15]. The research related to the progress of 5G communication systems and URLLC requirements is studied in the paper [16]; also, this paper examines tactile internet (TI). To achieve high reliability, a long code word with redundancy is required, which increases the latency, and a short packet is mandated to achieve low latency, which reduces reliability performance [17]. As a result, Shannon Capacity cannot be used, because, in that case, the code words must be extended, which is not applicable in URLLC. If Shannon's capacity is used in URLLC performance analysis, reliability and latency are underestimated [18]. In [19], the authors investigate a URLLC-based V2V communication scenario to minimize the decoding-error probability in the finite blocklength. The problem is formulated as a non-convex problem and the BCD algorithm is used to solve that problem. In [20], the joint optimization of blocklength and the unmanned aerial vehicle's (UAV) location to minimize the decoding error probability is studied, and a new algorithm is presented to solve the problem.

According to this description, works done in this paper are summarized as follows:

We first consider the problem of maximizing the weighted sum throughput (WST). This optimization problem is non-convex; after performing the necessary simplifications, this problem becomes two problems of convex optimization. Then we solve them by using the BCD algorithm and gradient ascent method (instead of the SCA algorithm), and in the derivative part of the gradient ascent algorithm, we use estimation methods.

Then we consider the problem of minimizing the total transmit power (TTP). This problem is also non-convex; after performing the necessary simplifications, this problem becomes a convex optimization problem, and to solve it, we repeat the gradient ascent method, and again in the derivative part of the gradient ascent algorithm, we use estimation methods.

Finally, according to the simulation results, the performance of the proposed method is better in most cases. In general, considering that the proposed method is less complicated, the slight performance deterioration in some cases can be tolerated and will be acceptable.

## II. SYSTEM MODEL AND FORMULAS

Here, we consider a downlink IoT communication system in Fig. 1, where data is transmitted wirelessly from an access point to k different devices. There is an eavesdropper whose purpose is to obtain transmitted data. We assume that all devices, access points, and eavesdropper have an antenna. Data transfer is done at the same time and at different frequencies.

Here, the frequency band is divided into several basic bandwidth units, denoted by $B_0$, and the total frequency bandwidth allocated to the kth device is denoted as $B_k = n_k B_0$. Where $n_k$ indicates the number of bandwidth units assigned to the kth device.

The channel coherence bandwidth is larger than the total bandwidth allocated to all devices [17]:

$$W_c = n_{max} B_0 \qquad (1)$$

$$\sum_{k=1}^{K} n_k \leq n_{max} \qquad (2)$$

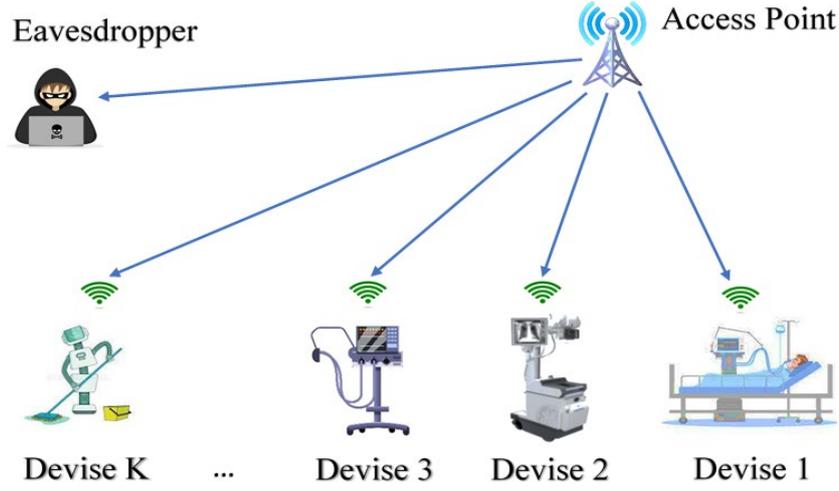

Fig. 1. IoT communication system

In (1), $n_{max}$ is the result of dividing $W_c$ by $B_0$. The number of channel uses obtained from the $N_k = B_K T$ equation [22]. In this equation, T represents the transmission time, which should be much less than the channel coherence time. Channels from the access point to the devices and the eavesdropper are indicated by $h_k^d$ and $h^e$, respectively, and these channels remain constant in each transmission.

The received signal-to-noise ratio (SNR) of the kth device is equal to [17]:

$$\gamma_k^d = \frac{p_k g_k^d}{n_k} \qquad (3)$$

where $p_k$ is the transmit power for the kth device, and the value of $g_k^d$ is obtained from the following equation, where $\sigma_{d,k}^2$ shows the noise power spectrum density at the kth device.

$$g_k^d = \frac{\left|h_k^d\right|^2}{\sigma_{d,k}^2 B_0} \qquad (4)$$

If the eavesdropper can access all the frequency bands, the equations listed above for the devices, also are valid for the eavesdropper.

$$\gamma_k^e = \frac{p_k g^e}{n_k} \qquad (5)$$

$$g^e = \frac{|h^e|^2}{\sigma_e^2 B_0} \qquad (6)$$

Another assumption of this paper is that all channel state information is available on the transmitter. However, the transmission of short packets is not safe from a non-zero decoding error probability and data leakage.

A lower bound on the maximum rate of secrecy communication rate is approximated as follows [17]:

$$r_k = C_k - \sqrt{\frac{V_k^d}{N_k}} \frac{Q^{-1}(\epsilon_k)}{\ln 2} - \sqrt{\frac{V_k^e}{N_k}} \frac{Q^{-1}(\delta_k)}{\ln 2} \qquad (7)$$

In this equation, $C_k$, represents maximum secrecy capacity, which is defined as follows:

$$C_k = \log_2(1 + \gamma_k^d) - \log_2(1 + \gamma_k^e) \qquad (8)$$

The channel dispersion and the Q-function are defined as follows:

$$V_k^x = 1 - (1 + \gamma_k^x)^{-2} ; \qquad x \in \{d, e\} \qquad (9)$$

$$Q(x) = \int_x^\infty \frac{1}{\sqrt{2\pi}} e^{\frac{-t^2}{2}} dt \qquad (10)$$

The total number of transferable bits for the kth device is as follows:

$$R_k = n_k B_0 T \left( C_k - \sqrt{\frac{V_k^d}{n_k B_0 T}} \frac{Q^{-1}(\epsilon_k)}{\ln 2} - \sqrt{\frac{V_k^e}{n_k B_0 T}} \frac{Q^{-1}(\delta_k)}{\ln 2} \right) \qquad (11)$$

The necessary condition to ensure a positive data rate is that $\gamma_k^d > \gamma_k^e$, [17], [9]. According to Equations (3) and (5), it can be concluded that $g_k^d > g^e$.

Our goal in this paper is to optimize the number of bandwidth units jointly, and power allocation by maximizing the weighted sum throughput and minimizing the total transmit power, respectively.

III. MAXIMIZING THE WEIGHTED SUM THROUGHPUT

As previously mentioned, our goal in this section is to optimize power allocation and bandwidth units to maximize the weighted throughput of all devices. Which is formulated as follows [17]:

$$(\text{Prob. 1}): \max_{p,n} \sum_{k=1}^K W_k R_k \qquad (12a)$$

$$s.t. \ \sum_{k=1}^K p_k \leq P_{max} \qquad (12b)$$

$$\sum_{k=1}^K n_k \leq n_{max} \qquad (12c)$$

$$p_k \geq 0, \quad \forall k = 1, \dots, k \qquad (12d)$$

$$n_k \in N^+, \quad \forall k = 1, \dots, k \qquad (12e)$$

In the above formula, $W_k$ is a positive coefficient, and $N^+$ represents a set of positive integers. Also $p = \{p_1, \dots, p_k\}$ and $n = \{n_1, \dots, n_k\}$.

To solve the problem, we relax the integer n to continuous variables and rewrite the formula (Prob.1) as follows:

$$(\text{Prob. 2}): \max_{p,n} \sum_{k=1}^K W_k R_k \qquad (13a)$$

$$s.t. \ (12b), (12c), (12d) \qquad (13b)$$

$$n_k \geq 0, \quad \forall k = 1, \ldots, K \tag{13c}$$

Because n and p are combined, the problem (Prob.2) is also difficult to solve. To make it solvable, one variable can be optimized, and the other kept constant

$$(\text{Prob. 2a}): \max_{p} \sum_{k=1}^{K} W_k R_k(p_k) \quad s.t. \ (12b), (12d)$$

$$(\text{Prob. 2b}): \max_{n} \sum_{k=1}^{K} W_k R_k(n_k) \quad s.t. \ (12c), (13c)$$

*A. Solution of (Prob.2a)*
Now, we solve the power allocation of Problem (Prob.2a) with given n. For this purpose, we define the following equations [17]:

$$\bar{g}_k^e \triangleq \frac{g^e}{n_k} \tag{14},$$

$$\bar{g}_k^d \triangleq \frac{g_k^d}{n_k} \tag{15},$$

$$L_k^d = \frac{Q^{-1}(\epsilon_k)\sqrt{N_k}}{\ln 2} \tag{16},$$

$$L_k^e = \frac{Q^{-1}(\delta_k)\sqrt{N_k}}{\ln 2} \tag{17}.$$

The value of $R_k(p_k)$ is rewritten as follows:

$$R_k(p_k) = \underbrace{N_k \log_2(1 + p_k \bar{g}_k^d) - N_k \log_2(1 + p_k \bar{g}_k^e)}_{f_k(p_k)} - \underbrace{\left(\sqrt{V_k^d}\, L_k^d + \sqrt{V_k^e} L_k^e\right)}_{y_k(p_k)} \tag{18}$$

According to the above equations, if $g_k^d$ is greater than $g^e$ ( $g_k^d > g^e$ ), we have $\bar{g}_k^d > \bar{g}_k^e$.
As shown in Appendix (1), both the functions $f_k(p_k)$ and $y_k(p_k)$ are concave, but their difference, $R_k(p_k)$, is not concave, because of that problem (Prob.2a) is a non-convex optimization problem, and it's difficult to solve.
The solution of p in the $(i-1)$th iteration is shown as $p^{(i-1)}$. In the following equation, $\chi_k(p_k^{(i-1)})$ represents the first-order derivative of $y_k(p_k)$ at $p_k^{(i-1)}$.

$$\chi_k\left(p_k^{(i-1)}\right) = \frac{\left(1+p_k^{(i-1)}\bar{g}_k^d\right)^{-3}\bar{g}_k^d L_k^d}{\left(1-\left(1+p_k^{(i-1)}\bar{g}_k^d\right)^{-2}\right)^{\frac{1}{2}}} + \frac{\left(1+p_k^{(i-1)}\bar{g}_k^e\right)^{-3}\bar{g}_k^e L_k^e}{\left(1-\left(1+p_k^{(i-1)}\bar{g}_k^e\right)^{-2}\right)^{\frac{1}{2}}} > 0 \tag{19}$$

By replacing $y_k(p_k)$ on the right-hand side of the (19), the optimization problem can be rewritten as follows, which is a convex optimization problem [17].

$$(\text{Prob. 2a} - 1): \max_{p} \sum_{k=1}^{K}\left(w_k f_k(p_k) - w_k \chi_k\left(p_k^{(i-1)}\right) p_k\right) \tag{20a}$$

$$s.t. \quad (12b), (12e) \tag{20b}$$

## B. Solution of (Prob.2b)

After solving section A, our goal is to optimize the number of bandwidth units of Problem (Prob.2b) with a given power allocation. For this purpose, we define the following equations [17]:

$$\bar{g}_k^d = p_k g_k^d \tag{21}$$

$$\bar{g}_k^e = p_k g_k^e \tag{22}$$

$$\tilde{N}_0 = B_0 T \tag{23}$$

$$\bar{L}_k^e = \sqrt{\tilde{N}_0} \frac{Q^{-1}(\sigma_k)}{\ln 2} \tag{24}$$

$$\bar{L}_k^d = \sqrt{\tilde{N}_0} \frac{Q^{-1}(\epsilon_k)}{\ln 2} \tag{25}$$

The value of $R_k(n_k)$ is rewritten as follows:

$$R_k(n_k) = \underbrace{\tilde{N}_0 n_k \log_2(1 + \frac{\bar{g}_k^d}{n_k}) - \tilde{N}_0 n_k \log_2\left(1 + \frac{\bar{g}_k^e}{n_k}\right)}_{F_k(n_k)} - \underbrace{(\sqrt{z_k^d(n_k)}\bar{L}_k^d + \sqrt{z_k^e(n_k)}\bar{L}_k^e)}_{Y_k(n_k)} \tag{26}$$

$$z_k^x(n_k) = n_k - \frac{n_k^3}{(n_k + \bar{g}_k^x)^2}, \quad x \in \{d, e\} \tag{27}$$

As shown in Appendix (2), both the functions $F_k(n_k)$ and $Y_k(n_k)$ are concave, but their difference, $R_k(n_k)$, is not concave; because of that, problem (Prob.2b) is a non-convex optimization problem, and it isn't easy to solve.

In the following equation, $\xi_k(n_k^{(j-1)})$ represents the first-order derivative of $Y_k(n_k)$ at $n_k = n_k^{(j-1)}$.

$$\xi_k\left(n_k^{(j-1)}\right) = \frac{\bar{L}_k^d\left(3(\bar{g}_k^d)^2 n_k^{(j-1)} + (\bar{g}_k^d)^3\right)}{2\sqrt{z_k^d(n_k^{(j-1)})}(n_k^{(j-1)} + \bar{g}_k^d)^3} + \frac{\bar{L}_k^e\left(3(\bar{g}_k^e)^2 n_k^{(j-1)} + (\bar{g}_k^e)^3\right)}{2\sqrt{z_k^e(n_k^{(j-1)})}(n_k^{(j-1)} + \bar{g}_k^e)^3} \tag{28}$$

By replacing $Y_k(n_k)$ on the right-hand side of the (28), the optimization problem can be rewritten as follows, which is a convex optimization problem [17].

$$(\text{Prob. 2b} - 1): \max_n \sum_{k=1}^K (w_k F_k(n_k) - w_k \xi_k\left(n_k^{(j-1)}\right) n_k) \tag{29a}$$

$$s.t. \quad (12c), (13c), \tag{29b}$$

## C. Algorithm analysis

Based on the analysis performed, we summarize the BCD algorithm as follows; in this algorithm, $R(n,p)$ is the weighted throughput defined as $R(n,p) = \sum_{k=1}^{K} W_k R_k(n,p)$.

To solve both (Prob. 2a − 1) and (Prob. 2b − 1) problems, we use the gradient ascent algorithm; $p^*$, and $n^*$ are the optimal solutions obtained by solving the problems (Prob. 2a − 1) and (Prob. 2b − 1). In this algorithm we have $R(n^{(t)}, p^{(t)}) \geq R(n^{(t-1)}, p^{(t-1)})$. This equation indicates that the solutions obtained by the BCD algorithm are increasing.

---

**Algorithm :** The BCD Algorithm for maximizing the Weighted Sum Throughput [17], [21]

---

1 **Initialize** $n = n^{(0)}$, $p = p^{(0)}$, accuracy $\varepsilon$, the iteration number $t = 1$. and calculate $R(n^{(0)}, p^{(0)})$ ;
2    **Repeat**
3      Set $n = n^{(t-1)}$, $i = 1$ ;
4      **Repeat**
5        Given $p^{(i-1)}$, find $p^{(i)}$ by solving problem (Prob.2a-1)(By useing gradient ascent algorithm and
6          estimation methods)
         and $i \leftarrow i + 1$ ;

7      **Until** $p$ Convergence ;
8      Update $p^{(t)} = p^*$ ;
9      Set $p = p^{(t)}$, $j = 1$ ;

10      **Repeat**
11        Given $n^{(j-1)}$, find $n^{(j)}$ by solving problem (Prob.2b-1)(By useing gradient ascent algorithm and
6          estimation methods)
         and $j \leftarrow j + 1$ ;

12      **Until** $n$ Convergence ;
13      Update $n^{(t)} = n^*$ and set $t \leftarrow t + 1$ ;
14    **Until**
$|R(n^{(t)}, p^{(t)}) - R(n^{(t-1)}, p^{(t-1)})| / R(n^{(t-1)}, p^{(t-1)}) \leq \varepsilon$

---

## IV. MINIMIZING THE TOTAL TRANSMIT POWER

Here, we minimize the total transmit power by jointly optimizing bandwidth and power allocation. Which is formulated as follows [17]:

$$(\text{Prob. 3}): \min_{p,n} \sum_{k=1}^{K} p_k \qquad (30a)$$

$$s.t. \ R_k \geq D_k^{min} \quad \forall k, \tag{30b}$$

$$(12c), (12d), (12e) \tag{30c}$$

$D_k^{min}$ represents the minimum power that the kth device needs. This problem is difficult to solve, due to the integer condition in the number of bandwidth units. For this reason, we define the following equations [17]:

$$V_k^x \approx 1, \tag{31}$$

$$x \in \{d, e\}, \tag{32}$$

$$\tilde{h}_k^d = T \frac{|h_k|^2}{\sigma_{d,k}^2}, \tag{33}$$

$$\tilde{h}^e = T \frac{\|h_e\|_2^2}{\sigma_e^2}, \tag{34}$$

$$\tilde{h}_k^d > \tilde{h}^e, \tag{35}$$

$$R_k \cong \tilde{R}_k = N_k \left( \log_2 \left( 1 + \frac{p_k \tilde{h}_k^d}{N_k} \right) - \log_2 \left( 1 + \frac{p_k \tilde{h}^e}{N_k} \right) \right) - N_k \left( \sqrt{\frac{1}{N_k}} \frac{Q^{-1}(\epsilon_k)}{\ln 2} - \sqrt{\frac{1}{N_k}} \frac{Q^{-1}(\delta_k)}{\ln 2} \right). \tag{36}$$

We can rewrite the problem (Prob. 3) as follows:

$$(\text{Prob. 4}): \min_{p,N} \sum_{k=1}^{K} p_k \tag{37a}$$

$$s.t. \ \tilde{R}_k \geq D_k^{min} \ , \forall k, \tag{37b}$$

$$\sum_{k=1}^{K} N_k \leq W_c T \tag{37c}$$

$$N_k \in \{B_0 T, 2B_0 T, \dots, n_{max} BT\} \ , \forall k, \tag{34d}$$

$$p_k \geq 0, \ \forall k \tag{37e}$$

Due to the discrete constraint on $N_k$, problem (Prob. 4) is difficult to solve. Because of that, we write it as (Prob. 4 − a).

$$(\text{Prob. 4} - a): \min_{p,N} \sum_{k=1}^{K} p_k \tag{38a}$$

$$s.t. \ (23b), (23c), (23e) \tag{38b}$$

$$N_k \geq 0 \tag{38c}$$

## V. ESTIMATION METHODS

Gradient descent and gradient ascent are the most important iterative algorithms in machine learning. Which is used in this paper to solve problems (Prob. 2a – 1), (Prob. 2b – 1) and (Prob. 4 – a). This algorithm is expressed as follows:

$$y_{i+1} = y_i - \sigma \nabla F(y_i) \quad (39)$$

To solve the derivative part of this algorithm, we use three estimation methods, Heun, Ralston, and forward-backward.

### A. Euler:

One of the most important methods of solving equations in numerical analysis is the use of Runge-Kutta methods. One of these methods, the first-order Runge kutta, is also known as the "Euler method", which is more popular. Our goal is to calculate the value of $\nabla F(y_i)$, from equation (36) with the Rang-Kutta methods.

$$\frac{dy}{dx} = f(x_i, y_i) \quad , \quad y_{i+1} = y_i + h\, f(x_i, y_i) \quad (40)$$

### B. Heun:

Different methods have been proposed for the second-order Range Kutta method, such as the Heun method, the middle point method, and the Ralston method. In this section, we will explain the Heun method.

$$y_{i+1} = y_i + h(\tfrac{1}{2}f(x_i, y_i) + \tfrac{1}{2}f(x_{i+1}, y_{i+1})) \quad (41)$$

### C. Ralston:

$$k_1 = f(x_i, y_i) \quad , \quad k_2 = f(x_i + \tfrac{3}{4}h, y_i + \tfrac{3}{4}k_1 h) \quad (42)$$

$$y_{i+1} = y_i + h(\tfrac{1}{3}k_1 + \tfrac{2}{3}k_2) \quad (43)$$

### D. Forward-Backward-Central difference:

In numerical analysis, there is another method for solving the differential equations called the forward-backward-central difference method.

According to the curve depicted in Fig. 2, if we assume that the point A is fixed and keep moving point B to the left, the forward derivative at point A is obtained. If we assume B is fixed and move point A, the backward derivative at point B is obtained and, if we assume the midpoint of the distance between A and B to be constant and bring A and B closer to the center, the central derivative is obtained at the midpoint between A and B.

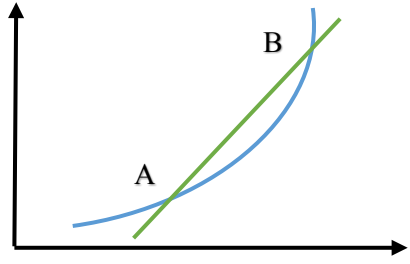

Fig. 2. A plot for explaining the forward-backward-central difference method

Forward difference: $\quad f'(y_i) = \frac{f(y_{i+1})-f(y_i)}{\Delta x}$ (44)

Backward difference: $\quad f'(y_i) = \frac{f(y_i)-f(y_{i-1})}{\Delta x}$ (45)

Central difference: $\quad f'(y_i) = \frac{f(y_{i+1})-f(y_{i-1})}{2\Delta x}$ (46)

$y_{i+1} = y_i \pm h \left(\frac{f(y_{i+1})-f(y_{i-1})}{2\Delta x}\right)$ (47)

## VI. RESULTS

In this section, we prepare the simulation results to evaluate the performance of the BCD algorithm by estimation methods. The adopted simulation parameters are given in table 1.

### A. Maximizing the weighted sum throughput

In this section, we have presented the simulation results obtained from $(\text{Prob. 2a}-1)$ and $(\text{Prob. 2b}-1)$, and have evaluated their performance with the estimation methods.

In Fig. 3, the convergence behavior of different methods are plotted. We used K=6 in this simulation. As can be seen from Fig. 3, all methods are converged after 20 iterations. Also, note that the computational complexity of the proposed estimation methods are lower than that of the reference method for each iterations. This means that the total simulation run time of the proposed methods are lower compared to that of the reference method.

Fig. 4 shows the simulation results of maximizing the weighted sum throughput (WST) versus the total power. In all four estimation methods, the WST increases as the total power increases. This behavior is expected, since by increasing the total power, the SNR increases, and so the weighted sum throughput increases. However, the heun method works better than the other methods, and the other methods are slightly different in performance.

Table 1. The values of the parameters used

| Parameters | symbol | value |
|---|---|---|
| bandwidth of channel unit | $B_0$ | 1 kHz |
| noise power spectrum density | $\sigma^2_{e,k}$ | -173 dBm/Hz |
| decoding error probability | $\varepsilon_k$ | $10^{-9}$ |
| information leakage | $\delta_k$ | $10^{-2}$ |
| period of time | T | 10 ms |
| channel coherence bandwidth | $W_c$ | 0.5 MHz |
| channel path loss | PL | $35.3 + 37.6 \, (\log_{10} l)$ [25] |
| The amount of distance between the device or eavesdropper and access point | $l$ | 200 m |

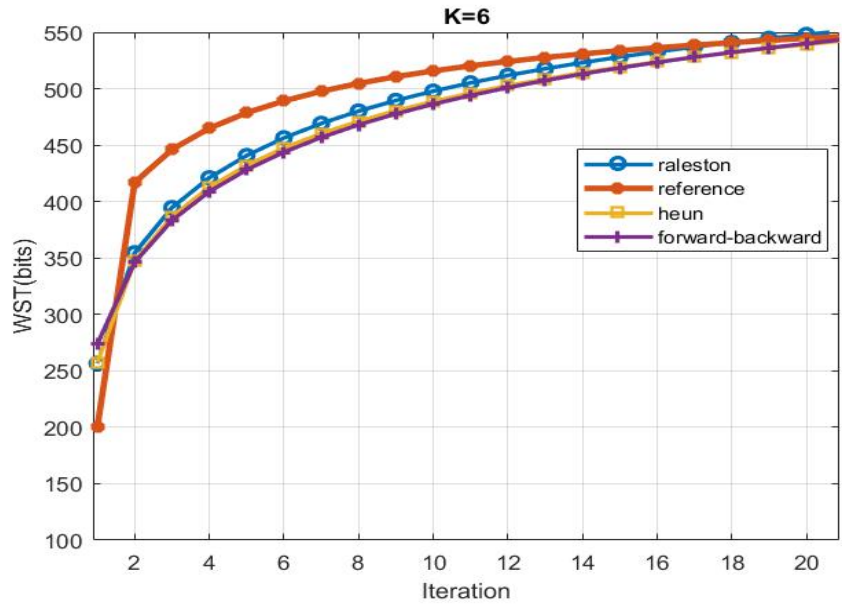

Fig. 3. The Convergence behavior of different methods for K=6.

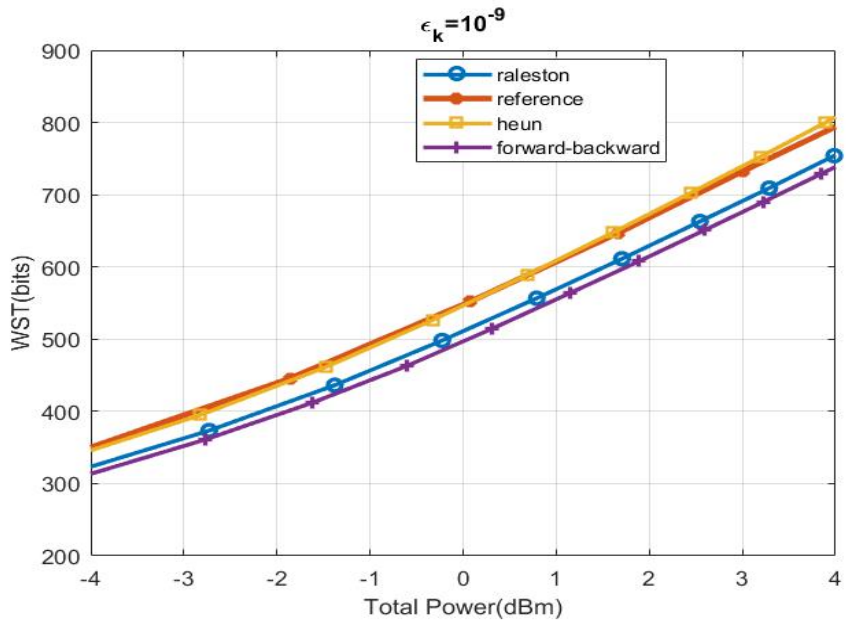

Fig. 4. The WST versus the total power of different methods.

Fig. 5, shows the WST versus the number of devices of different methods. In all four estimation methods, the WST increases as the number of devices increases. This again follows our expectations since by increasing the number of devices, the coherence bandwidth ($W_c$) also increases, and the increase in $w_c$ has a great effect on the increase of the WST. In this case the reference method works better than other methods; However, as the estimation methods are used in the proposed method with lower computational complexity compared to the reference method, the slight performance loss will be acceptable at the gain of simplicity of the proposed methods..

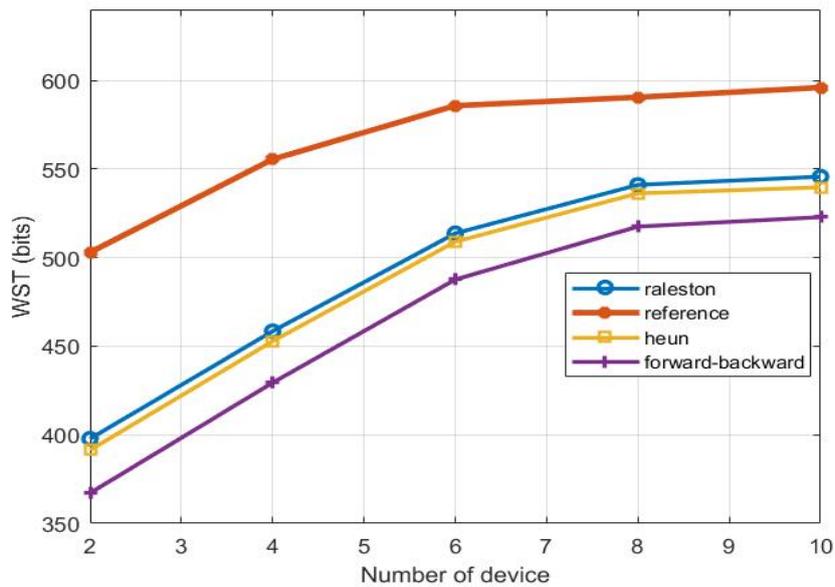

Fig. 5. WST vs the number of devices.

## B. Minimizing the total Transmit Power

In this section, we have presented the simulation results obtained from (Prob.4-a) and evaluated their performance with the estimation methods.

In Fig. 6, we investigate the effect of the packet size on total transmit power (TTP). As can be seen in this figure, the TTP increases in all methods as the packet size increases. This is due to the fact that the larger the size of the packets, the more power is required to achieve a successful sending rate. The four different method have the same performance, again with lower complexity of the proposed methods.

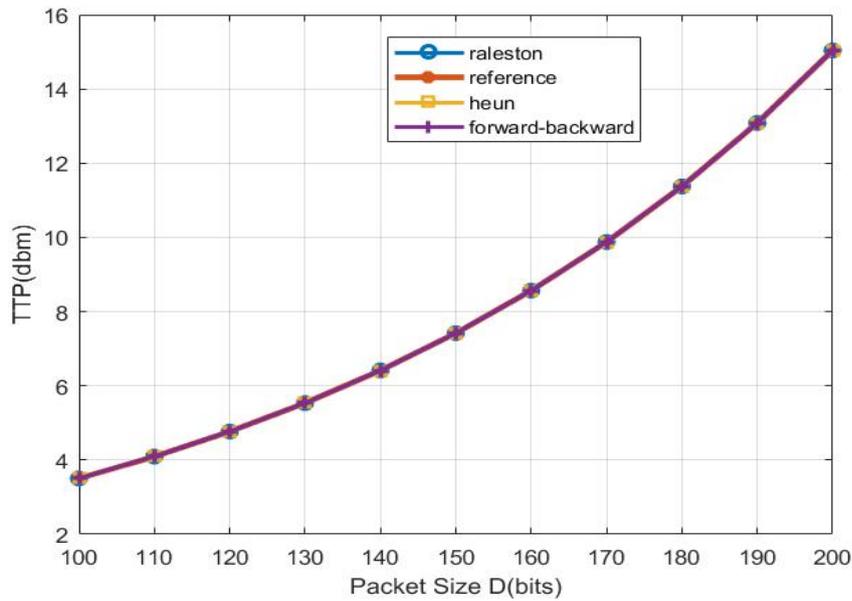

Fig. 6. TTP vs the packet size D

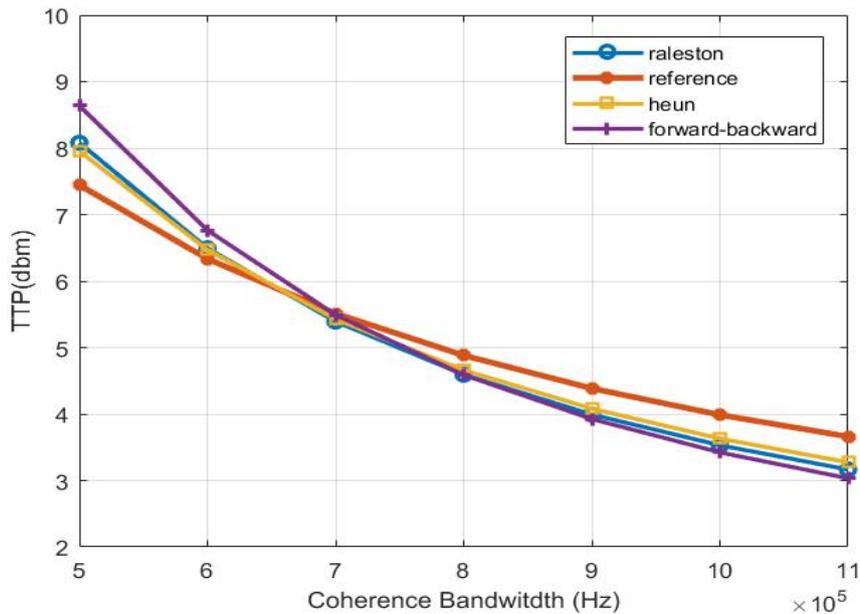

Fig. 7. TTP vs the channel coherence bandwidth

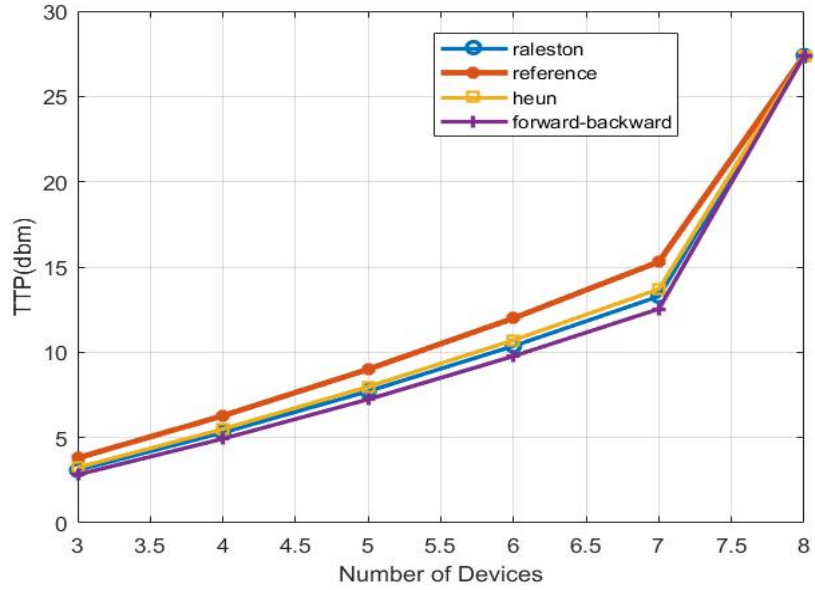

Fig. 8. TTP vs the number of devices.

In Fig. 7, we study the effect of channel coherence bandwidth on the performance of the TTP. In all estimation methods, the TTP decreases as the coherence bandwidth increases; in $W_c \geq 7 \times 10^5$, all methods work better than the reference method. This is due to the fact that by increasing the $W_c$, less power is required for the successful transmission. Also, for $W_c \geq 7 \times 10^5$, three estimation methods have almost the same performance.

Fig. 8, shows the TTP versus the number of devices. In all estimation methods, the TTP increases as the number of devices increases. The reason of this behaviour is that by increasing the number of devices, the number of bandwidth units allocated to each device decreases, and as a result, the more power is required for a successful transmission rate. In this figure, the three estimation methods work better than the reference method, and the Forward-Backward method presents the best performance among the other methods. As a numerical example, for seven devices, the total transmit power required in the Forward-Backward method is about 2.5 dB lower than that of the reference method.

## VII. CONCLUSION

In this paper, a downlink communication system is considered that AP transmits data to devices, and there is an eavesdropper whose purpose is to obtain transmitted data. Under this context, the problem of maximizing the weighted sum throughput and the problem of minimizing the total transmit power, are considered. Both of the WST and TTP problems are non-convex. Then the necessary simplifications are performed to convert the problem of non-convex optimization to the problem of convex optimization. To solve the problem of maximizing the weighted sum throughput, the BCD and gradient ascent algorithm are used, and to solve the problem of minimizing the total transmit power, the gradient ascent algorithm is used; also, estimation methods are used to solve the derivative part of the gradient ascent algorithm. The simulation results show that the performance of the proposed method is faster than the reference method in some cases, and in some cases the proposed method is better than the reference method, and also the proposed method has less computational complexity.

## APPENDIX 1

In this section, we proved that $f_k(p_k)$ and $y_k(p_k)$ are concave functions. The second derivative of $f_k(p_k)$ and $y_k(p_k)$ are as follows equation, Since $\bar{g}_k^d > \bar{g}_k^e$, $f_k''(p_k)$ will be less than zero.

$$f_k''(p_k) = \frac{N_k}{\ln 2} \frac{\bar{g}_k^e - \bar{g}_k^d}{(1+p_k\bar{g}_k^d)(1+p_k\bar{g}_k^e)} \left( \frac{\bar{g}_k^e}{1+p_k\bar{g}_k^e} + \frac{\bar{g}_k^d}{1+p_k\bar{g}_k^d} \right) < 0$$

$$y_k''(p_k) = -\frac{(\bar{g}_k^d)^2 L_k^d \left(3 - 2(1+p_k\bar{g}_k^d)^{-2}\right)}{\left(1-(1+p_k\bar{g}_k^d)^{-2}\right)^{\frac{3}{2}}(1+p_k\bar{g}_k^d)} - \frac{(\bar{g}_k^e)^2 L_k^e \left(3 - 2(1+p_k\bar{g}_k^e)^{-2}\right)}{\left(1-(1+p_k\bar{g}_k^e)^{-2}\right)^{\frac{3}{2}}(1+p_k\bar{g}_k^e)} < 0$$

According to the equation (18), $R_k(p_k)$ is the difference of two concave function, but $R_k(p_k)$ is not a concave function.

## APPENDIX 2

In this section, we proved that $F_k(n_k)$ and $Y_k(n_k)$ are concave functions. The second derivative of $F_k(n_k)$ and $Y_k(n_k)$ are as follows equation. Since $\bar{g}_k^d > \bar{g}_k^e$, $f_k''(p_k)$ will be less than zero.

$$F_k''(n_k) = \frac{\tilde{N}_0(\bar{g}_k^e - \bar{g}_k^d)\left((\bar{g}_k^e + \bar{g}_k^d)n_k + 2\bar{g}_k^e \bar{g}_k^d\right)}{\ln 2 (n_k + \bar{g}_k^e)^2 (n_k + \bar{g}_k^d)^2} < 0$$

$$Y_k''(n_k) = \frac{2\frac{\partial^2 z_k^d(n_k)}{\partial n_k^2} z_k^d(n_k) - \left(\frac{\partial z_k^d(n_k)}{\partial n_k}\right)^2}{4 z_k^d(n_k)\sqrt{z_k^d(n_k)}} \tilde{L}_k^d + \frac{2\frac{\partial^2 z_k^e(n_k)}{\partial n_k^2} z_k^e(n_k) - \left(\frac{\partial z_k^e(n_k)}{\partial n_k}\right)^2}{4 z_k^e(n_k)\sqrt{z_k^e(n_k)}} \tilde{L}_k^e$$

By replacing (24) into the following equation, the value of $Y_k''(n_k)$ will be smaller than zero.

Where $\frac{\partial^2 z_k^x(n_k)}{\partial n_k^2}$ is given by:

$$\frac{\partial^2 z_k^x(n_k)}{\partial n_k^2} = -\frac{6n_k(\bar{g}_d^x)^2}{(n_k + \bar{g}_d^x)^4} < 0, \qquad x \in \{d, e\}.$$

According to the equation (23), the $R_k(n_k)$ is the difference of two concave function; however, the $R_k(n_k)$ is not a concave function.